\documentclass[a4paper,nofootinbib,showpacs,preprintnumbers,floatfix,superscriptaddress,pra,twocolumn]{revtex4-1}
\usepackage{amsmath, amsthm, amssymb}
\usepackage{graphicx}
\usepackage{dcolumn}
\usepackage{bm}
\usepackage{color}
\usepackage{amsmath}
\usepackage{amsfonts}
\usepackage{amssymb}
\providecommand{\U}[1]{\protect\rule{.1in}{.1in}}
\newcommand{\sbullet}{\,\begin{picture}(1,1)(-0.5,-2)\circle*{3}\end{picture}\,}

\newcommand{\ket}[1]{| #1 \rangle}
\newcommand{\bra}[1]{\langle #1 |}

\newcommand{\tr}{\mathrm{Tr}}

\newcommand{\proj}[1]{\ensuremath{|#1\rangle \langle #1|}}
\newcommand{\beq}{\begin{equation}}
\newcommand{\eeq}{\end{equation}}
\newcommand{\bea}[1]{\begin{equation}\begin{array}{#1}}
\newcommand{\eea}{\end{array}\end{equation}}
\newcommand{\beqn}{\begin{eqnarray}}
\newcommand{\eeqn}{\end{eqnarray}}
\providecommand{\openone}{\mathbbm{1}}
\renewcommand{\rho}{\varrho}

\renewcommand{\c}[1]{\mathcal{#1}}

\begin{document}
\title{Noisy One-Way Quantum Computations: The Role of Correlations}
\author{Rafael Chaves}
\email{rafael.chaves@icfo.es} \affiliation{ICFO-Institut de
Ciencies Fotoniques, Mediterranean Technology Park, 08860
Castelldefels (Barcelona), Espa\~na} \affiliation{Instituto de
F\'{\i}sica, Universidade Federal do Rio de Janeiro, Rio de
Janeiro, Brasil} \affiliation{Physikalisches Institut der
Albert--Ludwigs--Universit\"at, Freiburg, Deutschland}
\author{Fernando de Melo}
\email{fnando.demelo@gmail.com}
\affiliation{Instituut voor Theoretische Fysica, Katholieke Universiteit Leuven, Leuven, Belgi\"e}
\affiliation{Physikalisches Institut der Albert--Ludwigs--Universit\"at, Freiburg, Deutschland}

\date{\today}

\begin{abstract}
A scheme to evaluate computation fidelities within the one-way
model is developed and explored to understand the role of
correlations in the quality of noisy quantum computations. The
formalism is promptly applied to many computation instances, and
unveils that a higher amount of entanglement in the noisy resource
state does not necessarily imply a better computation.
\end{abstract}

\pacs{03.67.-a, 03.67.Mn, 42.50.-p}
\maketitle

%%% extra new command -- do not move it to the preamble!
\renewcommand{\i}{{\rm i}}

%%%%%%%%%%%%%%%%%%%%%%%%
%%  Introduction %%%%%%%
%%%%%%%%%%%%%%%%%%%%%%%%
\section{Introduction}
\label{introduction}Since the advent of quantum
computation~\cite{qc}, entanglement, a clear-cut quantum
mechanical feature, is commonly believed the key resource
behind it. Not surprisingly, the importance of correlations for
quantum computations has been a much debated subject. For
\emph{pure} state quantum computations certainly some entanglement
is necessary if the quantum protocol is not to be efficiently
simulated by classical means~\cite{Jozsa2003, Vidal2003}. However,
entanglement is only necessary, but not a sufficient condition for
an exponential gain of quantum computations over classical ones.
There are quantum protocols that despite of producing highly
entangled states can still be efficiently simulated
classically~\cite{Jozsa2003, KnillGottesman1999}.

The scenario for \emph{mixed} state quantum computations is far more
subtle~\cite{Jozsa2003}. A model for mixed state quantum computation
introduced in~\cite{Knill1998}, in which the input state consists of
a single qubit in a pure state and all the others in a uniform
incoherent sum of classical alternatives -- and therefore not
entangled--, offers an exponential speed-up to problems
that are believed intractable by classical
computers~\cite{DattaThesis}. Also, room temperature NMR
implementations of quantum information tasks~\cite{Jones2001}, which
employ a rather noisy state where entanglement is known not to be
present~\cite{Braunstein98}, seem to still present gain over
classical computations~\cite{classical}. Nevertheless, in these
cases, generation of entanglement during the computation itself
cannot be ruled out~\cite{DattaThesis}. A definitive statement about
the influence of entanglement is thus challenging.

A clean investigation of the entanglement role in noisy quantum
computations is however possible within the  One-Way
model~\cite{oneway}. In this model, local (projective)
measurements on a highly entangled resource state are responsible
for input preparation, the required computation and final
read-out. No entanglement is created during the computation. We
have thus a clear distinction between the entanglement creation
and its use as a resource.

Employing this model of computation we address here still another
facet of the entanglement role in noisy quantum computations: How
does the noise affecting the entangled resource state impact on
the ``quality" (fidelity) of the computation? Does a more
entangled resource state always empower better computations? Or in
more practical terms: should one always try to minimize the
influence of the environment over the entanglement such as to
maximize the fidelity of a computation? To answer in the
\emph{negative} to these questions, we derive an expression for
the fidelity of any one-way computation when the resource state
undergoes various types of decoherence. Our results extend to
noisy computations the assertion~\cite{TooEntangled} that a high
entangled state is not always advantageous for a measurement based
quantum computation.

This article is organized as follows: In Sec. \ref{onewaymodel},
we briefly review the one-way model for quantum computations. In
Sec.~\ref{onewaymodelunderdecoherence}, we discuss the effects of
various models of decoherence on one-way computations, and derive
the expression for the fidelity in such cases. In Sec.~\ref{apps}
we apply the developed formalism to various computation instances,
showing that a higher amount of entanglement in the noisy resource
state does not necessarily imply  better computations. In
particular we obtain that some instances of the
Deutsch-Jozsa~\cite{DJ92} algorithm, which in the circuit model of
quantum computation generically create entangled
states~\cite{EntangledDJ1998}, within the one-way framework
require \emph{no entanglement} for their execution. In
Sec.~\ref{ancilla} we analyze the effects of decoherence in the
ancilla-driven quantum computation proposed in Ref.
\cite{ancilladriven}. In Sec.~\ref{conclusion} we summarize our
results, and draw some conclusions.

%%%%%%%%%%%%%%%%%%%%%%%%
%% Oneway Model %%%%%%%%
%%%%%%%%%%%%%%%%%%%%%%%%

\section{One-way model of quantum computation}
\label{onewaymodel}In the one-way model~\cite{oneway} all the
interactions between qubits, and local unitary transformations
needed by the protocol are exchanged by a prior entanglement in a
graph-state, and the possibility to make adaptive local
measurements. A graph-state is defined by a set of vertices
$\mathcal{V}$, and a collection of edges $\mathcal{E}$. In each
vertex $i$ sits a qubit ($\mathcal{H}_2$) initialized in a state
$|+_{i} \rangle=(|0_{i}\rangle+|1_{i}\rangle)/\sqrt{2}$, and an
edge represents an interaction between  vertices $\{i,j\}$ given
by $CZ_{ij}=\proj{0_i}\otimes\openone_j+\proj{1_i}\otimes Z_j$.
Hereafter $\left\{ \openone,X,Y,Z \right\}$ will represent the
usual $\left\{ \sigma_{0}, \sigma_{1},\sigma_{2},\sigma_{3}
\right\} $ Pauli matrices. The $N$-qubit graph state is then
\begin{equation}
|{G_{(\mathcal{V},\mathcal{E})}}\rangle=\prod_{\{i,j\}\in\mathcal{E}%
}CZ_{ij}|+_{k}\rangle^{\otimes k\in\mathcal{V}}.
\end{equation}
Starting with a $N$-qubit graph state, an one-way computation is
carried out by measuring $M$ qubits, and the remaining $N-M$
qubits encode the protocol answer. The algorithm is then defined
by a triple $\left\{\theta_{i},\alpha_{i},s_{i}\right\}$, with
instructions on the measurement basis
$\ket{M_{k_{i}}^{s_i}(\theta_{i},\alpha_{i})}$ for the
$i$-th qubit. Here
\begin{align}
\left\vert M_{0_{i}}^{s_i}\left( \theta_{i},\alpha_{i}\right)
\right\rangle  &
=\cos\frac{\alpha_{i}}{2}\left\vert 0\right\rangle +\sin \frac{\alpha_{i}}{2} \; e^{-\i(-1)^{s_{i}%
}\theta_{i}}\left\vert 1\right\rangle \label{base}\\
\left\vert M_{1_{i}}^{s_i}\left( \theta_{i},\alpha_{i}\right)
\right\rangle  &
=\sin\frac{\alpha_{i}}{2}\left\vert 0\right\rangle -\cos\frac{\alpha_{i}}{2} \; e^{-\i(-1)^{s_{i}%
}\theta_{i}}\left\vert 1\right\rangle \nonumber
\end{align}
with $s_{i}\equiv s_{i}(\vec{k})=s_{i}(k_{1,}\ldots,k_{M}) \in
\{0,1\}$ the adaptation parameter that depends on the outcome
$k_{j} \in \{0,1\}$ of previous measurements and implicity on the
algorithm being considered (hereafter, the notation $\vec{x}$
represents the M-tuple $x_1,\ldots,x_M$). The need for adaptations
stems from the requirement of turning every computation
deterministic, despite of the intrinsic randomness associated with
each quantum measurement. Adaptations introduce a temporal order
for the measurements, and thus classical correlations others than
the already present in the initial state. The only two instances
of $\alpha_i$ necessary for all computations are: $\alpha_{i}=0$
for measurements along the $z$ direction, and $\alpha_{i}=\pi/2$
for measurements in $x$-$y$ plane (equator) of the Bloch sphere.
The measurements are non-adaptive when the basis are given by the
Pauli operators $\left\{X,Y,Z\right\}$ eigenvectors, and possibly
adaptive otherwise.  Lastly, the desired answer is given aside
some local unitary transformations, final by-products ($BP$) of
the computation, determined by the classical outcomes $\vec{k}$ of
all measured qubits. These by-products can be dealt with by
classical post-processing.

Defined a computation to be performed on a graph state
$|{G_{(\mathcal{V},\mathcal{E})}}\rangle$, the $M$ qubits to be
measured can be immediately written in their measurement basis
(adaptations included) what explicit the answer:
\begin{equation}
\ket{G_{(\mathcal{V},\mathcal{E})}}%
 = \frac{1}{2^\frac{M}{2}}\sum_{\vec{k}}\bigotimes_{i=1}^{M}\ket{M_{k_{i}}^{s_i}
 (\theta_{i},\alpha_{i})}\ket{A_{\vec{k}}(\vec{\theta},\vec{\alpha})} ,\label{statemeasurement}%
\end{equation}
with $\ket{A_{\vec{k}}(\vec{\theta},\vec{\alpha})} =
BP_{\vec{k}} \ket{A_{\vec{0}}(\vec{\theta},\vec{\alpha})}$, and $\ket{A_{\vec{0}}(\vec{\theta},\vec{\alpha})}$
the desired answer without the by-products. The latter, in turn,
can be expressed as local unitaries \beq
BP_{\vec{k}}={\bigotimes_{i=N-M+1}^{N}} (-1)^{f_{i,Sig}(\vec{k})}
X_{i}^{f_{i,X}(\vec{k})}Z_{i}^{f_{i,Z}(\vec{k})}, \label{BP} \eeq
 with $f_{Sig}$, $f_X$ and $f_Z$ boolean functions of the outcomes $\vec{k}$
defined by the protocol at hands. In a noise free computation, the
sign boolean function $f_{Sig}$ only introduces a global phase on
each answer, and it is then of no importance. However, under the
action of the environment, some of the answers will be mixed, and
this phase turns a relative one. As such, it cannot be neglected.
It is important to notice that out of $2^{M}$ possible measurement
outcomes in a given protocol, only $4^{N-M}$ of them possibly lead
to different answers (modulo a global phase), meaning that in
general many of answers are the same.

%%%%%%%%%%%%%%%%%%%%%%%%%%%%%%%%%%%%%%%
%% Fidelity formula: oneway  %%%%%%%%%%
%%%%%%%%%%%%%%%%%%%%%%%%%%%%%%%%%%%%%%%
\section{Fidelity of noisy one-way quantum computations \label{onewaymodelunderdecoherence}}

Once hardware (graph state) and
algorithm (measurement basis + possible adaptations) are defined,
we want to gauge how does the quality of a computation decay due to
different noisy environments, and how  this is related to the decay
of the initial entanglement resource.

The standard measure for the computation quality is the output
fidelity $\mathcal{F}$~\cite{qc, RemStatePrep, SingletFraction,
Weinstein, Chaves, Tomoyuki, Tomoyuki2}. This measure compares the
desired state, $\ket{\Psi_\text{out}}$, with the actually obtained
one, $\rho_\text{out} $, returning  ${\mathcal
F}(\ket{\Psi_\text{out}},\varrho_\text{out}) :=
\bra{\Psi_\text{out}} \varrho_\text{out} \ket{\Psi_\text{out}}$.
The fidelity measure is also used to define the error threshold for
quantum computations~\cite{preskill}.

In what follows, we consider that each qubit of the initial graph state is
individually coupled to its own environment. A great variety
of single qubit open dynamics are encompassed by the map
\begin{equation}
\Lambda_k(\sbullet)= \sum_{j=0}^{3}
\lambda_{j}^k(t)\sigma_{j}\,\sbullet\,\sigma_{j}+\mu^k(t)\left[
\begin{array}[c]{c}%
\sigma_{3}\,\sbullet\,\sigma_{0}+\sigma_{0}\,\sbullet\,\sigma_{3}\\
-\i\sigma_{1}\,\sbullet\,\sigma_{2}+\i\sigma_{2}\,\sbullet\,\sigma_{1}%
\end{array}
\right]  \label{generalchannel}%
\end{equation}
with $\lambda_{0}^k(t)=\left(1+2e^{-C_k t}+e^{-B_k t}\right)/4$,
$\lambda_{1}^k(t)=\lambda_{2}^k(t)=\left(1-e^{-B_kt}\right)/4$,
$\lambda_{3}^k(t)=\left(1-2e^{-C_k t}+e^{-B_k t}\right)/4$, and
$\mu^k(t)=(2S_k-1)\left(1-e^{-B_k t}\right)/4  $. For the $k$-th qubit, the parameters $B_k$ and
$C_k$ are, respectively, the decay rate of inversion and
polarization, and $S_k\in\left[0,1\right]$ depends on the
temperature of the bath~\cite{dur05}.

The decohered state after time $t$ is obtained by the composition
of the individual evolutions $\Lambda_k$, i.e., $\varrho(t) = \Lambda(\proj{G_{(\mathcal{V},\mathcal{E})}}):=
\Lambda_{1}\otimes ...\otimes
\Lambda_{N}(\proj{G_{(\mathcal{V},\mathcal{E})}})$. The assumption
of mutually independent environments is well justified whenever
the separation between the vertices is large enough so that
collective effects need not be taken into account.

To evaluate the computation fidelity after the noisy evolution, we
note that, once defined a measurement basis, only the diagonal
terms in such basis are relevant to the measurement outcome. The
action of the channel (\ref{generalchannel}) on a general
measurement basis term $\left\vert M_{k_{i}}\right\rangle
\left\langle M_{k_{i}^\prime}\right\vert$ is such that non-diagonal ($k_i \neq k^\prime_i$) terms
are mapped onto non-diagonal terms, while diagonal ($k_i = k^\prime_i$) terms evolve as
\begin{align}
\left\vert M_{k_{i}}\right\rangle \left\langle
M_{k_{i}}\right\vert \mapsto\left(  1-p_i\right)  \left\vert
M_{k_{i}}\right\rangle \left\langle M_{k_{i}}\right\vert
+p_i\left\vert M_{k_{i}\oplus1}\right\rangle \left\langle
M_{k_{i}\oplus1}\right\vert \nonumber\\
+ \text{off diagonal terms};
\label{decoonbasis}
\end{align}
where henceforth we use a simplified notation whenever ambiguities are not possible.

In the equation above, $p_i$ is  a function of time, and of the parameters
describing the channel (\ref{generalchannel}). In the case of  a
measurement in $x$-$y$ plane
\begin{equation}
p_i^{xy}=\lambda_{1}^i(t)+\lambda_{3}^i(t), \label{decoonbasisxy}
\end{equation}
and for a measurement on the $z$ direction it reads
\begin{equation}
p_i^z=2\lambda_{1}^i(t)+(-1)^{k_{i}+1}2\mu^i(t).
\label{decoonbasisz}
\end{equation}

As an example, consider the following two particular noise instances (which will be used later):\\
$i)$ phase-flip error (pf) -- with a probability $p_\text{pf}/2$
the state $\ket{0}+\ket{1}$ is mapped onto $\ket{0}-\ket{1}$ and
vice versa. This is obtained by setting $B_i=0$ and
$C_i=2\Gamma_\text{pf}$ in (\ref{generalchannel}). Accordingly,
the state evolution maps $\rho_i \mapsto (1-p_\text{pf}/2)\rho_i +
p_\text{pf} (Z \rho_i Z)/2$, with $p_\text{pf} =
[1-\exp(-2\Gamma_\text{pf}\, t)]$.  The impact of the decoherence
on the measurement basis is then given by \eqref{decoonbasisxy}
and \eqref{decoonbasisz} %%
\begin{align}
p_{i,\text{pf}}^{xy} &= p_\text{pf}/2,\nonumber \\
p_{i,\text{pf}}^{z}&=0. \label{dephonbasis}
\end{align}
$ii)$ white noise (w) -- add to the previous case the
possibility of errors into the other independent directions $x$
and $y$. This amounts to exchange the initial state with a
maximally mixed one with probability $p_\text{w}$. This is
described by setting $S_i=1/2$ and $B_i=C_i=4\Gamma_\text{w}$ in
(\ref{generalchannel}). Under this dynamics, the state evolves to
$(1-p_\text{w})\rho_i + p_\text{w} \openone/2$, with $p_\text{w} = 1
- \exp(-4 \Gamma_\text{w} t)$. In this case, we have
\begin{align}
p_{i,\text{w}}^{xy}=p_{i,\text{w}}^{z}=p_\text{w}/2.
\label{whiteonbasis}
\end{align}

It is important to notice that depending on the required
measurement by an algorithm, some noisy maps might have no effect, and the corresponding  measurement outcomes are
thus not disturbed. This feature will be further exploited in
section~\ref{without_adaptations}.

Note also that these
decoherence processes can be interpreted as if the performed
measurement was not perfect, being unable to distinguish between
the two possible outcomes with probability $p_i$.

\bigskip

Now we are set to evaluate the fidelity of \emph{any} one-way
noisy quantum computation. Given a result $\vec{r}$ for the
measurements, we want to determine
$\mathcal{F}(\ket{A_{\vec{r}}(\vec{\theta},\vec{\alpha})},
\rho_{\vec{r}})$, with $\rho_{\vec{r}}$ the $(N-M)$ qubit state
encoding the noisy protocol answer. We are thus
interested on the projection of
$\Lambda(\proj{G_{(\mathcal{V},\mathcal{E})}})$ onto
$\bigotimes_{i=1}^M \proj{M_{r_i}^{s_i(\vec{r})}}$.
\begin{widetext}
However, since $\bra{M_{r_i}^{s_i(\vec{r})}} M_{k_i}^{s_i(\vec{k})}\rangle \neq 0$ in general, we first note that
\begin{equation}
\ket{M_{k_{i}}^{s_i(\vec{k})}} = \frac{1}{2}
\Big\{[1+(-1)^{k_{i}} e^{-2\i \theta_i .\left(  s_i(\vec{k})\oplus
s_i(\vec{r})\right) }]\ket{M_{0_{i}}^{s_i(\vec{r})}} +
[1-(-1)^{k_{i}} e^{-2\i \theta_i .  \left( s_i(\vec{k})\oplus
s_i(\vec{r})\right)}]\ket{M_{1_{i}}^{s_i(\vec{r})}}\Big\},
\end{equation}
 and therefore, the state in \eqref{statemeasurement} can be rewritten as:
\beq \ket{G} = \frac{1}{2^\frac{3
M}{2}}\sum_{\vec{k}}\bigotimes_{i=1}^{M}\left\{ [1+(-1)^{k_{i}}
e^{-2\i \theta_i . \left( s_i(\vec{k})\oplus
s_i(\vec{r})\right)}]\ket{M_{0_{i}}^{s_i(\vec{r})}} +[1-(-1)^{k_{i}}
e^{-2\i \theta_i . \left(  s_i(\vec{k})\oplus
s_i(\vec{r})\right)}]\ket{M_{1_{i}}^{s_i(\vec{r})}}
\right\}\ket{A_{\vec{k}}}.
\end{equation}
In the last two expressions above we used that when the
measure is in the $z$ direction there is no adaptation, i.e.,
$s_i=0$.

Due to the noise, the components
$\proj{M_{0_i}^{s_i(\vec{r})}}$ and
$\proj{M_{1_i}^{s_i(\vec{r})}}$  mix according to the prescription
in Eq.\eqref{decoonbasis}, reducing the fidelity of the protocol.
The state after the action of the noise reads: \beq
\frac{1}{2^{3M}}\sum_{\vec{k},\vec{l}}\bigotimes_{i=1}^M\\
\left\{%
\begin{matrix}
[1+(-1)^{k_{i}} e^{-2\i \theta_i . \left(  s_i(\vec{k})\oplus
s_i(\vec{r})\right)}][1+(-1)^{l_{i}} e^{2\i \theta_i .\left(
s_i(\vec{l})\oplus s_i(\vec{r})\right) }]
\begin{bmatrix}
(1-p_i)\proj{M_{0_i}^{s_i(\vec{r})}}\\
 + p_i \proj{M_{1_i}^{s_i(\vec{r})}}
\end{bmatrix}
 \\
+ [1-(-1)^{k_{i}} e^{-2\i \theta_i . \left(  s_i(\vec{k})\oplus
s_i(\vec{r})\right)}] [1-(-1)^{l_{i}} e^{2\i \theta_i . \left(
s_i(\vec{l})\oplus s_i(\vec{r})\right)}]
\begin{bmatrix}
(1-p_i)\proj{M_{1_i}^{s_i(\vec{r})}}\\ + p_i
\proj{M_{0_i}^{s_i(\vec{r})}}
\end{bmatrix}
\\
+ \text{ off diagonal terms}
\end{matrix}%
\right\}\Lambda(\ket{A_{\vec{k}}}\bra{A_{\vec{l}}}).
\eeq

The projection onto the subspace corresponding to $\vec{r}$ leads
the remaining $N-M$ qubits in the state: \beq \rho_{\vec{r}} =
\frac{1}{2^{3M}}\frac{1}{\mathcal{Z}_{\vec{r}}}\sum_{\vec{k},\vec{l}}\prod_i^M
\left\{%
\begin{matrix}
(1-p_i)[1+(-1)^{k_{i}+r_{i}} e^{-2\i \theta_i . \left(  s_i(\vec{k})\oplus s_i(\vec{r})\right)}] [1+(-1)^{l_{i}+r_{i}} e^{2\i \theta_i . \left(  s_i(\vec{l})\oplus s_i(\vec{r})\right)}]\\
+ p_i [1-(-1)^{k_{i}+r_{i}} e^{-2\i \theta_i . \left(
s_i(\vec{k})\oplus s_i(\vec{r})\right)}][1-(-1)^{l_{i}+r_{i}}
e^{2\i \theta_i . \left(  s_i(\vec{l})\oplus s_i(\vec{r})\right)}]
\end{matrix}%
\right\}\Lambda(\ket{A_{\vec{k}}}\bra{A_{\vec{l}}}); \eeq with
$\mathcal{Z}_{\vec{r}}$ the probability of obtaining the outcome
$\vec{r}$. Note that, in the above expression, the channels acting
in each qubit can be different, and might influence the state for
different time intervals. This is an important feature, for the
qubits might be of different nature, and could be measured
at different times.

We are thus in  position to evaluate the computation fidelity
$\mathcal{F}_\text{OneWay}(\vec{r})=\bra{A_{\vec{r}}}\varrho_{\vec{r}}\ket{A_{\vec{r}}}$
for the decohered graph state, to get: \beq
\mathcal{F}_\text{OneWay}(\vec{r}) =
\frac{1}{2^{3M}}\frac{1}{\mathcal{Z}_{\vec{r}}}\sum_{\vec{k},\vec{l}}\prod_i^M
\left\{%
\begin{matrix}%
(1-p_i)[1+(-1)^{k_{i}+r_{i}} e^{-2\i \theta_i . \left(  s_i(\vec{k})\oplus s_i(\vec{r})\right)}] [1+(-1)^{l_{i}+r_{i}} e^{2\i \theta_i . \left(  s_i(\vec{l})\oplus s_i(\vec{r})\right)}]\\
+ p_i [1-(-1)^{k_{i}+r_{i}} e^{-2\i \theta_i . \left(
s_i(\vec{k})\oplus s_i(\vec{r})\right)}][1-(-1)^{l_{i}+r_{i}}
e^{2\i \theta_i . \left( s_i(\vec{l})\oplus s_i(\vec{r})\right)}]
\end{matrix}%
\right\}A_{\vec{r},\vec{k},\vec{l}}\label{fidelity}; \eeq with $
A_{\vec{r},\vec{k},\vec{l}}= \bra{A_{\vec{r}}} \Lambda
\big(\ket{A_{\vec{k}}} \bra{A_{\vec{l}}}\big)\ket{A_{\vec{r}}}$.
\end{widetext}
%%%
Instead of looking for the fidelity of a particular outcome
$\vec{r}$ of the computation, one is often more interested on the
average fidelity over all the outcomes, simply given by
%%%
\begin{equation}
\overline{\mathcal{F}}_{OneWay}=\sum_{\overrightarrow{r}}\mathcal{Z}_{\overrightarrow{r}%
}\mathcal{F}_{OneWay}\left(  \overrightarrow{r}\right).
\label{fidaverage}
\end{equation}
%%%%

With this expression in hands
and the results in~\cite{GraphStatesEntanglement}, which allow for
the evaluation of noisy graph-state entanglement, one can compare
the dynamics of entanglement in the state used as resource with
the fidelity dynamics of \emph{any} noisy one-way computation.
Furthermore, from the expression~(\ref{fidelity}) one can
immediately infer that the computation fidelity shows a continuous
decay in time -- as each $p_j$ is continuous function of time. That is
in clear contrast with a generic entanglement evolution, where the
amount of entanglement may vanish in finite time~\cite{terra01}.
As noted in~\cite{Weinstein} (considering the particular effects
of individual decoherence in a four-qubit cluster state), the
finite time disentanglement does not cause changes in the
behaviour of the computation fidelity. Entanglement decay is shown
here, in great generality, to not display a one-to-one correlation
with computation quality.

%%%%%%%%%%%%%%%%%%%%%%%%%%%%%%%%%%%%%%%
%%%%%%%%%%%%%%%% NA computations %%%%%%
%%%%%%%%%%%%%%%%%%%%%%%%%%%%%%%%%%%%%%%
\subsection{Computations without adaptations}
\label{without_adaptations}

Computations that need no adaptations (NA) turn out to be a very
interesting subset of all possible computations. If a computation
requires no adaptations, $s_i(\vec{k})=0 \; \forall i \text{ and }
\forall \vec{k}$,  then all the outcomes happen with the same
probability, $\mathcal{Z}_{\vec{r}}= 1/2^M \; \forall \vec{r}$. In
this case, Eq.~\eqref{fidelity} simplifies to: %%
\beq \mathcal{F}_\text{OneWay}^{\text{NA}}(\vec{r}) =
\sum_{\vec{k}}\prod_i^M (1-p_i)^{1\oplus k_i\oplus r_i}
p_i^{k_i\oplus r_i}
A_{\vec{r},\vec{k},\vec{k}}\;\label{fidelityNA}. \eeq %%
Expression~(\ref{fidelityNA})  shows that the overall effect of
quite general decoherence processes, as parametrized in
(\ref{generalchannel}), over the non-adaptative measured qubits is
to incoherently combine all the possible noisy answers associated
with such measurements. The sign boolean function $f_{Sig}$ in the
definition of the by-products~\eqref{BP} can, also in such cases, be
safely ignored.

Further simplification is possible for non-adaptative computations
when the channel acting on the answer-qubits is given by a Pauli
map $\Gamma$, obtained setting $\mu(t)$=0 in the general
expression (\ref{generalchannel}). In this case
$\Gamma[BP_{\vec{k}} \proj{A_{\vec{0}}} BP^\dagger_{\vec{k}}]=
BP_{\vec{k}} \Gamma[\proj{A_{\vec{0}}}] BP^\dagger_{\vec{k}}$ and
$A_{\vec{r},\vec{k},\vec{k}} =
A_{\vec{0},\vec{r}\oplus\vec{k},\vec{r}\oplus\vec{k}}$, which
implies that
$\mathcal{F}_\text{OneWay}^{\text{NA}}(\vec{r})=\mathcal{F}_\text{OneWay}^{\text{NA}}(\vec{0})$.
The average fidelity (\ref{fidaverage}) is then the same as the
fidelity of any outcome, tremendously simplifying its evaluation.

A greater insight on the nature of correlations necessary for an
one-way computation without adaptations is possible when
considering noise of the  form %%
\begin{equation}
\Lambda_{j}(\sbullet)=(1-p_{j})\,\sbullet+p_{j}
R_{\hat{n}_{j}}(\phi_{j})\,\sbullet\,
R_{\hat{n}_{j}}^{\dag}(\phi_{j}) \label{maps2fixedpoles}
\end{equation}
with $R_{\hat{n}_{j}} (\phi_{j})=\exp(-i
\phi_{j}\hat{n}_{j}\cdot\vec{\sigma}/2)$ a rotation around the
axis $\hat{n}_{j}$ in the Bloch sphere. This map has two invariant
 states, the eigenvectors $\ket{\hat{n}_{j\,0,1}(\phi_{j})}$
of the rotation. Now remember that a measurement in a certain
basis can be done directly on it, or by first applying a   unitary
transformation to a convenient basis and then the measurement. It
stands for a simple relabelling. Therefore, for all non-adaptive
measurements (NAMs), if the noise is of the type in (\ref{maps2fixedpoles}), it
is possible to find $U(\phi_j,\theta_j,\alpha_j)$ that transforms
$\ket{M_{k_{j}}(\theta_{j},\alpha_{j})}\rightarrow
\ket{\hat{n}_{j\,k_j}(\phi_{j})}$. This perfectly protects the
outcome probability distribution of the NAMs, even for highly
mixed resource states, as the measurement outcomes won't be
affected. For computations in which all the $N$ qubits are
measured in a non-adaptive fashion, this procedure protects the
whole computation, and the graph state can be  exchanged by a
mixed state with only classical correlations. The noisy
computation without adaptations can thus be classically simulated.

To exemplify this, consider the state in which the $i$-th qubit is
to be measured in the $X$ basis, namely:
\begin{equation}
\left\vert G_{N}\right\rangle =\frac{1}{\sqrt{2}}\left( \left\vert
+_i\right\rangle  \left\vert A_{+}\right\rangle +\left\vert
-_i\right\rangle \left\vert A_{-}\right\rangle \right),
\end{equation}
with $\ket{A_{\pm}} = (1/\sqrt{2})\left( \left\vert
G_{N-1}\right\rangle \pm  {\bigotimes_{j\in\mathcal{N}_{i}}}
Z_{j}\left\vert G_{N-1}\right\rangle \right)$, and $\mathcal{N}_i$ the neighbours of the $i$-th qubit.
 Applying a bit-flip channel $\Lambda_i^{X} (\sbullet) := (1-p)\sbullet + pX \sbullet X$ on the
$i$-th qubit decreases the entanglement between this qubit and the
rest of the graph, while it does not affect the measurement
outcome. Therefore, since applying $X$ on a qubit $i$ of a graph
is equivalent to apply $Z$'s in all its neighbors
$\mathcal{N}_{i}$~\cite{dur05}, preparing the state,
\begin{equation}
\begin{array}
[c]{c}%
\frac{1}{2} \big( \left\vert G_{N}\right\rangle \left\langle G_{N}
\right\vert +  \prod_{j\in\mathcal{N}_{i}} Z_{j} \left\vert
G_{N}\right\rangle
\left\langle G_{N}\right\vert \prod_{j\in\mathcal{N}_{i}} Z_{j} \big)\\
=\\
\frac{1}{2}\big(\proj{+_i}\otimes\proj{A_+} +
\proj{-_i}\otimes\proj{A_-}\big)
\end{array}
\end{equation}
%%%
which only bears \emph{classical correlations} between the $i$-th
qubit and the rest of the graph state, and is invariant under the
application of bit flip to the $i$-th qubit, leads to the same
probability distribution  of outcomes for a measurement of the
$i$-th qubit on the $X$ basis. Moreover it generates the same
answers $\left\{  \left\vert A_{\pm} \right\rangle \right\}$, and
this procedure can be iterated to the remaining qubits to be
measured non-adaptively.

The NAMs are related to the so-called Clifford-group
transformations part of a quantum protocol, and as such can be
simulated efficiently in a classical
computer~\cite{KnillGottesman1999, oneway, montanaro}. From the
framework here presented, it is clear that the entanglement
between the qubits to be measured non-adaptively and the rest of
the graph state can be interchanged by simple classical
correlations encoded in a mixed state without compromising the
computation. When adaptations are necessary this scheme
cannot prevail. Adaptive measurements are thus related to the
quantum part of the computation, where some resilient entanglement
may be of use.

%%%%%%%%%%%%%%%%%%%%%%%%%%%%%%%%%%%%%%%%%%%%%%%%%%%%%%%%%%%%%%%
%%%%%%%%%%%%%%%%%%%%%%% Apps %%%%%%%%%%%%%%%%%%%%%%%%%%%%%%%%
%%%%%%%%%%%%%%%%%%%%%%%%%%%%%%%%%%%%%%%%%%%%%%%%%%%%%%%

\section{Applications}
\label{apps}

In the following we apply the formulae developed above to specific
examples, and compare the fidelity dynamics of noisy one-way
quantum computations to the entanglement decay of the resource
state. In general, a mismatch between the two dynamics is
found:  higher entanglement is not connected with higher quality
computations.

%%%%%%%%%%%%%%%%%%%%%%%%%%%%%%%%%%%%%%%%%%
%%%%%%%%%%%%%%% RSP %%%%%%%%%%%%%%%%%%%%%%

\subsection{Remote State Preparation (RSP)}
\label{RSPqubits} Take the simplest possible one-way protocol,
i.e, to remotely prepare the single qubit state
$\cos(\phi/2)\ket{0} - \i\sin(\phi/2)
\ket{1}$~\cite{RemStatePrep}. Within the one-way model we start
with a graph state of two qubits, $\ket{G_2} = (\ket{00}  +
\ket{01} + \ket{10} - \ket{11})/2$, and apply a measurement on the
first qubit as to produce the desired state on the second one. For
this task, one chooses to measure the observable with eigenvectors
$\ket{M_k (\theta, \pi/2)} = \left[\ket{0}+(-1)^k \exp(- i
\theta)\ket{1} \right]/\sqrt{2}$, where $k\in\{0,1\}$ represents
the two possible measurement outcomes. After the measurement, the
second qubit is left in the state $
\ket{\Psi_\text{out}}=X^k\left[\cos(\phi/2) \ket{0} - \i
\sin(\phi/2) \ket{1}\right]/\sqrt{2}$. Apart from the by-product
$BP_k = X^k$, which is known, the desired state is obtained with
maximal fidelity. However, quantum computations are prone to
errors. Let's assume, for simplicity, a favourable scenario where
only the first qubit (the one to be measured) is subjected to
noise. The initial two-qubit state evolves then to
$(\Lambda_1\otimes\openone)(\proj{G_2})$. Since no adaptations are necessary, and the assumption that the second qubit is not under the action of an environment, the mean fidelity for this protocol, as evaluated by~\eqref{fidelityNA} reads:
\beq
\overline{\c{F}}_{\text{RSP}}  = \c{F}_{\text{RSP}}(0) = (1-p_1) A_{0,0,0} + p_1 A_{0,1,1}.
\eeq
For the measurement is on the $x$-$y$ plane then $p_1=p_1^{xy}$. Furthermore, $A_{0,0,0} = |\bra{A_0} A_0\rangle|^2=1$, and  $A_{0,1,1} = |\bra{A_0} X \ket{A_0}|^2=0$. This leads to $\c{F}_{\text{RSP}} = (1-p_1^{xy})$, which now depends only on the specific nature of the noise acting on the first qubit.

To address the
connection between computation fidelity and entanglement, consider
the two instances of open system dynamics introduced in
Sec.~\ref{onewaymodelunderdecoherence}:
%
%%%%%%%%%%%%
% figure 1
%%%%%%%%%%%%
\begin{figure} [t!]
\begin{center}
\includegraphics[width=\linewidth]{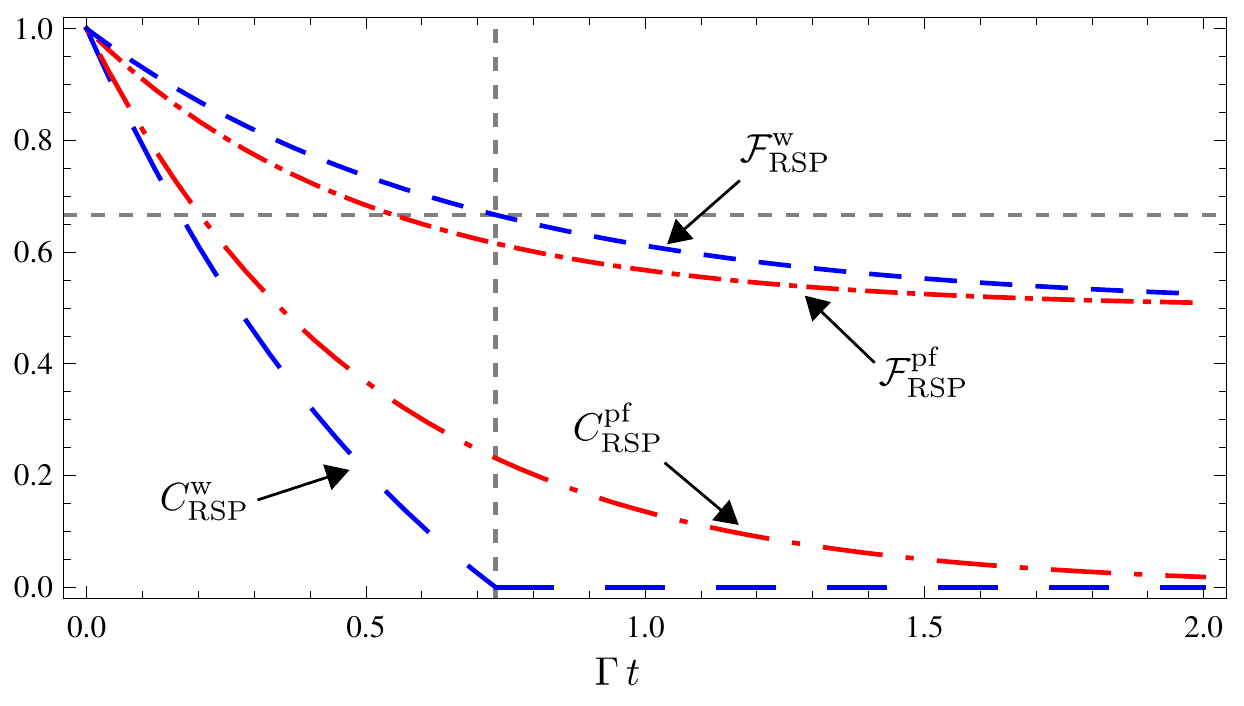}
\caption{(Color online) {\bf $RSP$: Entanglement decay vs. computation
fidelity} ($\Gamma_\text{w} = 0.375\,\Gamma_\text{pf} = \Gamma$).
A less entangled state may lead to a higher computation fidelity
(the same order is obtained if the
negativity~\cite{negativityvidal} is used instead of concurrence).
In the white noise case, entanglement vanishes when the fidelity
reaches 2/3 (horizontal dashed line)~\cite{nota}. Entanglement is
thus superfluous for achieving fidelities below this threshold.
\label{fig:Exampleentanglement}}
\end{center}
\end{figure}
%%%%%%%%%%
%%%%%%%%%%
%%%%%%%%%%%%
% figure 2
%%%%%%%%%%%%
\begin{figure} [t!]
\begin{center}
\includegraphics[width=\linewidth]{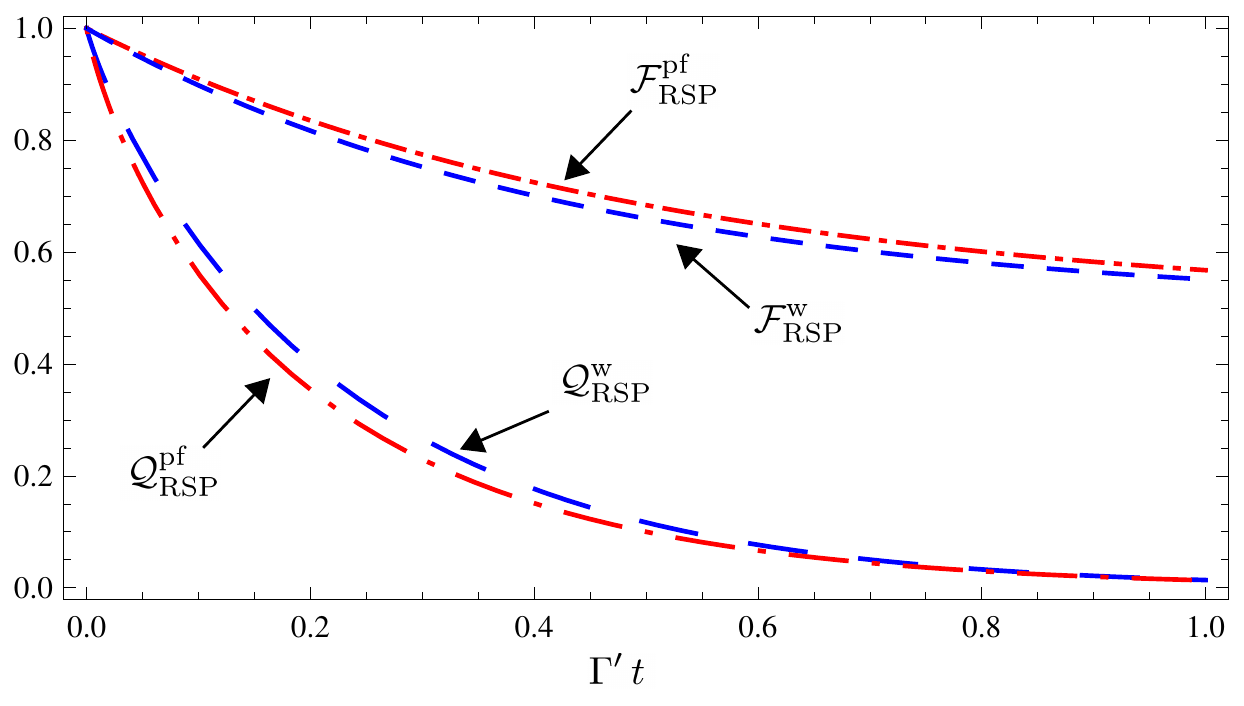}
\caption{(Color online) {\bf $RSP$: Quantum discord decay vs. computation
fidelity} ($\Gamma_\text{w} = 0.57 \,\Gamma_\text{pf} =
\Gamma^\prime$). As for concurrence, a state with less quantum
discord may lead to a better quality computation. Quantum discord
vanishes only asymptotically.\label{fig:counterExamplediscord}}
\end{center}
\end{figure}
%%%%%%%%%%%%%%%%
\\ \noindent $i)$ phase-flip error (pf) -- the state evolution maps
$\proj{G_2} \mapsto (1-p_\text{pf}/2)\proj{G_2} + p_\text{pf}/2(Z
\otimes \openone)\proj{G_2}(Z \otimes \openone)$, with
$p_\text{pf} = [1-\exp(-2\Gamma_\text{pf}\, t)]$. The entanglement
dynamics of the noisy resource state can be inferred by its
concurrence~\cite{wooters98}, $C_{\text{RSP}}^\text{pf}(t) =
\exp(-2\Gamma_\text{pf} t)$. Now, applying the state preparation
protocol described above to the decohered state, the output
fidelity, given that $p_1^{xy}= p_{\text{pf}}/2$, is ${\mathcal
F}_{\text{RSP}}^\text{pf}(t)= [1+\exp(-2\Gamma_\text{pf}\, t)]/2.$
The
correlation between decreasing entanglement with decreasing fidelity is as supposed. \\
$ii)$ white noise (w) -- by adding noise to the other independent directions, the resource state evolves to
$(1-p_\text{w})\proj{G_2} + p_\text{w} \openone/4$, with
$p_\text{w} = 1 - \exp(-4 \Gamma_\text{w} t)$. As before, we
evaluate the entanglement dynamics via concurrence,
$C_{\text{RSP}}^\text{w}(t) = \max\{0, [3\exp(-4\Gamma_\text{w}
t)-1]/2\}$. Finally, the protocol fidelity with white noise, $p_1^{xy}= p_{\text{w}}/2$, reads: ${\mathcal
F}_{\text{RSP}}^\text{w}(t)=[1+\exp(-4 \Gamma_\text{w} t)]/2$.
Once again, a smaller value of entanglement leads to a worst
computation.

Nevertheless, a comparison between both situations shows
unexpected behaviour (see Fig.~\ref{fig:Exampleentanglement}): the
computation fidelity is higher when the entanglement is more
fragile against disturbances. Note that even after the
entanglement is fully exhausted, the white noise case still
outperforms the always entangled phase-flip case. Even in the
simplest one-way protocol the entanglement is neither
\emph{sufficient} nor \emph{necessary} signature of higher quality
for the noisy quantum computation.

%%%%%%%%%%%%%%%%%%%%%%%%%%%%%%%%%%
%% counterexample: discord %%%%%%%
%%%%%%%%%%%%%%%%%%%%%%%%%%%%%%%%%%
In fact, this reasoning can be extended to quantum discord, a
recently proposed measure of quantum correlations which does not
include only entanglement~\cite{Ollivier2001}. This measure
attracted lots of attention lately, since it seems to pin-point
efficient quantum computations even in the apparent absence of
entanglement~\cite{DattaThesis}. Quantum discord is defined as
${\mathcal Q}(\varrho) = {\mathcal I}(\varrho) -{\mathcal
C}(\varrho)$, where the mutual quantum information: \beq {\mathcal
I}(\varrho) = S(\varrho_A) + S(\varrho_B) - S(\varrho) \eeq is a
measure of total correlations, and \beq {\mathcal C}(\varrho) =
\sup_{\{\Pi_k\}} S(\varrho_A) - S(\varrho_A^k | \{\Pi_B^k\}) \eeq
is a measure of classical correlations, with the supremum taken
over all sets of orthogonal projectors $\{\Pi_B^k\}$. As usual,
$S(\varrho) = -\tr (\varrho \log_2 \varrho)$ denotes the von
Neumann entropy, and $\varrho_i =
\tr_{j\neq i} (\varrho)$ with $i,j = A, B$ is the partial density
matrix. For classical systems $\mathcal{I}=\mathcal{C}$, thus
equivalent definitions for the mutual information, resulting  zero
quantum discord.

Evaluating the quantum discord, as shown in Ref.~\cite{luo08}, for
the toy-protocol described above shows that a lower quantum
discord may lead to a higher computation fidelity (see
Fig.~\ref{fig:counterExamplediscord}). Also quantum discord cannot
be employed to signal higher quality in a noisy quantum
computation scenario.

%%%%%%%%%%%%%%%%%%%%%%%%%%%%%%%%%%
%% example: REQ            %%%%%%%
%%%%%%%%%%%%%%%%%%%%%%%%%%%%%%%%%%

As a last possible signature of a higher quality noisy quantum
computation we may employ the measure of non-classicality very
recently proposed in~\cite{MEP}, namely the minimum entanglement
potential ($MEP$) of a given state $\rho_{A}$. As a result of the
following activation protocol (see~\cite{MEP} for details):
\begin{quotation}
$i)$ act with local unitaries $U_{A}$ in each subsystem of
$\rho_{A}$,

$ii)$ interact, through a $CNOT$, each subsystem of $\rho_{A}$ with
an ancilla initialized in the state $\ket{0}$,
\end{quotation}
any non-classical state becomes entangled (for any choice of
$U_{A}$) with the ancillary system $A'$. The minimum entanglement
generated across the $A : A'$ split quantifies the
non-classicality of state, that is,
\begin{equation}
MEP \left ( \rho_{A} \right ) = min_{U_{A}} E_{A : A'} \left (
\rho'_{A: A'} \right ),
\end{equation}
where $\rho'_{A: A'}$ is the system-ancilla state generated in the
end of the activation protocol.

Choosing as a measure of entanglement the negativity~\cite{negativityvidal}, we can
readily compute the $MEP$ for  the resource state in the two noise scenarios above mentioned. In these cases we get  $MEP\left(\Lambda^{\text{pf/w}}[\proj{G_2}]\right )= \left( 1-p^{\text{pf/w}} \right) =: MEP_{\text{RSP}}^{\text{pf/w}}$. For the RSP protocol then the relation $MEP_{\text{RSP}}^{\text{pf/w}} = 2 \c{F}_{\text{RSP}}^{\text{pf/w}} -1$ holds, and therefore a higher $MEP$ implies higher computation quality. At least in this simple example, the non-classicality of the resource state, quantified by the minimum entanglement potential, seems to be correlated to the quality of the computation.

%%%%%%%%%%%%%%%%% Primitives For Universal Quantum Computation%%%%%%%%%%%%%%%%%%%%%%%%%%

\subsection{Primitives for Universal Quantum Computation}
\label{example3qubits}

The basic gates for universal quantum computation are a generic
single qubit rotation ($R$), and a two-qubit controlled operation,
say, a controlled-not ($CNOT$) gate. How is the fidelity of these
basic building blocks related to the entanglement of their
resource states?

%%%%%%%%%%%%
% figure 3
%%%%%%%%%%%%
\begin{figure} [htf]
\begin{center}
\includegraphics[width=\linewidth]{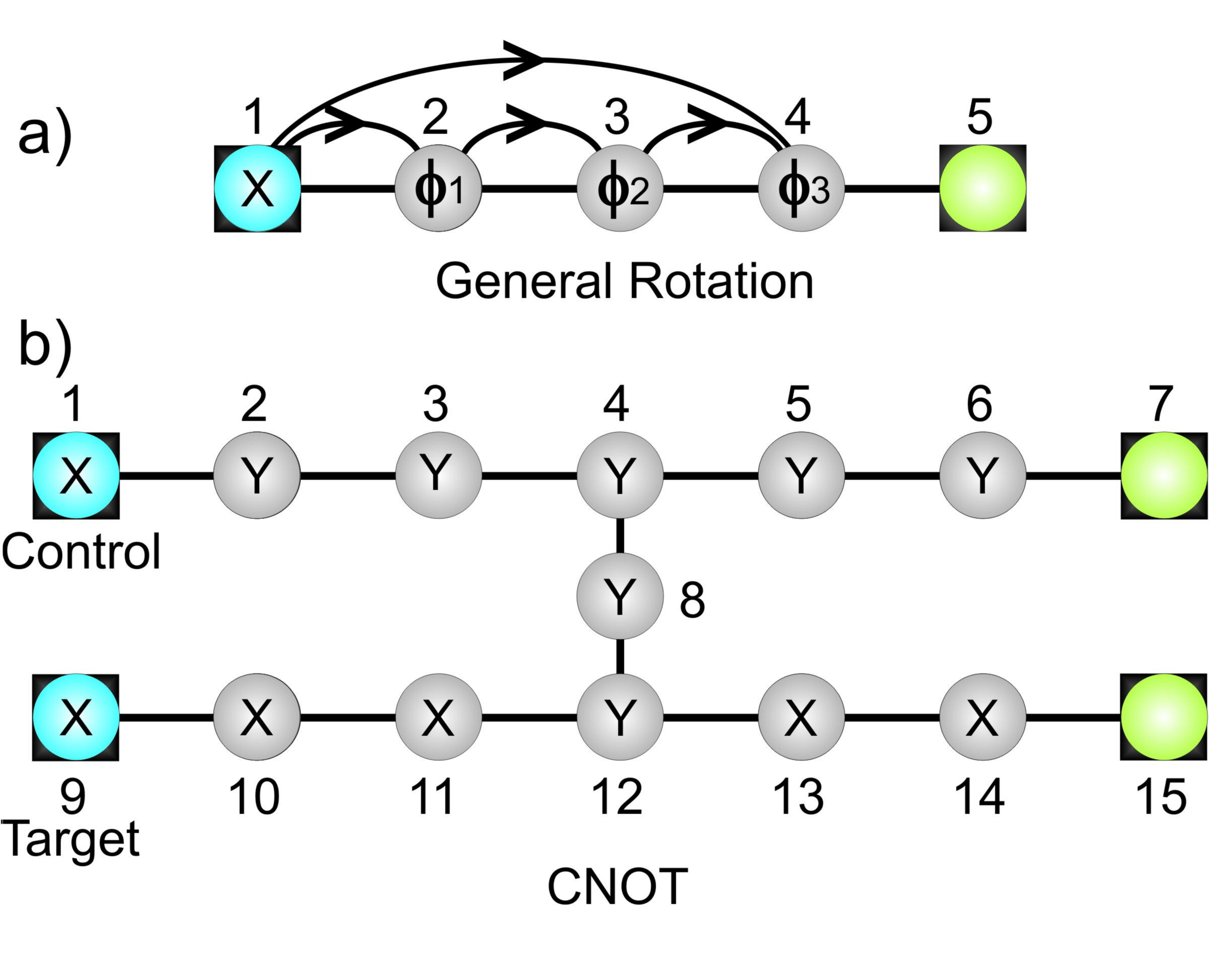}
\caption{(Color online) \textbf{Universal gates.} One-way
implementation of a universal set of gates. The blue qubits encode
the initial state, and the green ones the output state after the
measurements. Grey qubits represent intermediate steps of the
computation.  All the measurements required are in the $xy$ plane
($\alpha_i = \pi/2$), with the latitude angle $\theta_i$ specified
for each qubit in the picture.  \textbf{a) Single qubit rotation}.
A general rotation requires adaptive measurements, which are
indicated by the arrows above. These arrows are translated into
the adaptation parameter for each qubit as $s_1(\vec{k}) =  0,
s_2(\vec{k}) =  k_1, s_3(\vec{k}) =  k_2$, and  $s_4(\vec{k}) =
k_1+k_3$. \textbf{b) Controlled-not.} Despite of this be an
entangling gate, it requires no adaptations. This means that once
the input state is defined, it can be classically simulated (see
Section~\ref{without_adaptations}).} \label{fig:primitives}
\end{center}
\end{figure}
%%%%%%%%%%%%

%%%%%%%%%%%%
% figure 4
%%%%%%%%%%%%
\begin{figure} [htf]
\begin{center}
\includegraphics[width=\linewidth]{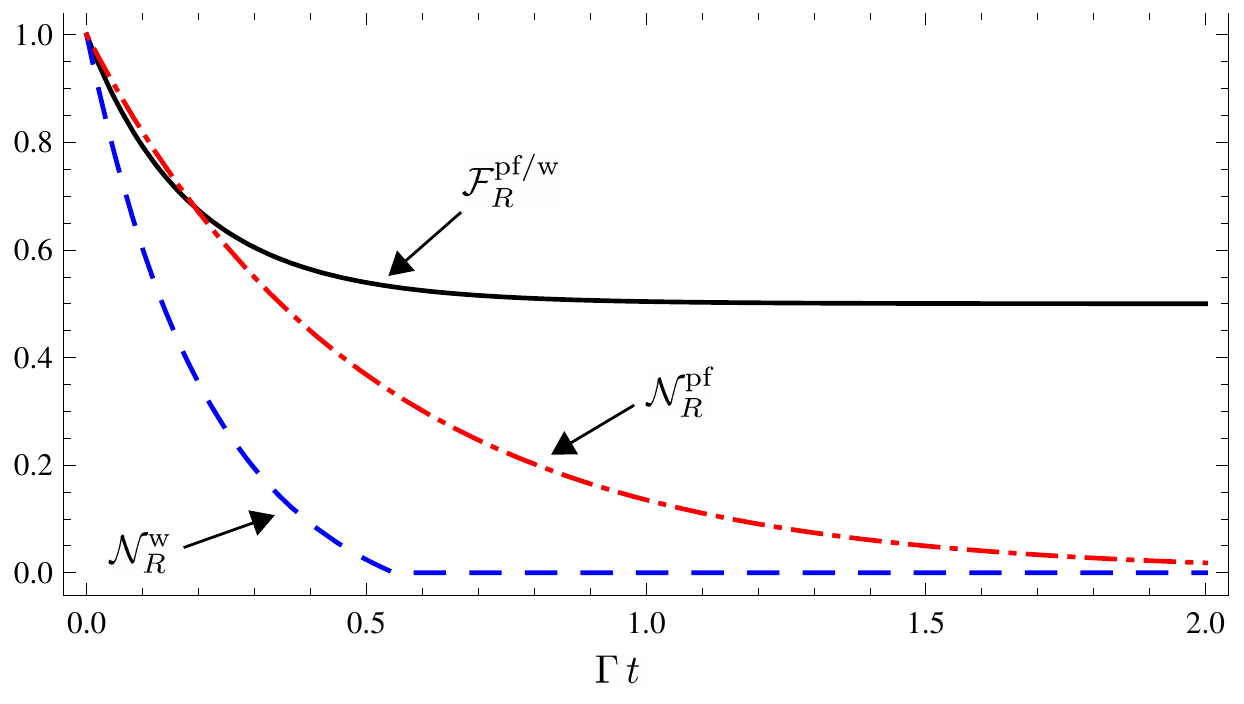}
\caption{(Color online) \textbf{$R$: Entanglement decay vs.
computation quality:} Noisy rotation of the initial state
$\ket{\psi_{\text{in}}}=\ket{0}$ setting
$\phi_1=\phi_2=\phi_3=\pi/4$, when  $p_{\text{w}}=p_{\text{pf}}$,
i.e., $\Gamma_{\text{pf}}= 2 \Gamma_{\text{w}} = \Gamma$. In this
case, $\mathcal{F}^{\text{pf}}_{R}=\mathcal{F}^{\text{w}}_{R}$,
represented  by the solid black line. Nevertheless, the
entanglement dynamics differs for each noisy instance. The
entanglement between the output qubit  with the remaining of the
cluster, as estimated by the negativity~\cite{negativityvidal}, is
depicted by the dashed blue line for the white noisy case
($\mathcal{N}_{R}^{\text{w}}$), and by the dot-dashed red line for
phase flip ($\mathcal{N}_{R}^{\text{pf}}$). A lower amount of
entanglement can thus yield a as good computation as a higher
amount. Similar result holds for the $CNOT$ gate.
\label{fig:fidelities_rotation_cnot}}
\end{center}
\end{figure}
%%%%%%%%%%%%
\medskip

Any $U(2)$ rotation can be decomposed into successive rotations
over three different angles, known as the Euler angles, as
$R(\phi_1,\phi_2,\phi_3)=R_{\hat{x}}(\phi_3) R_{\hat{z}}(\phi_2)
R_{\hat{x}}(\phi_1)$. Any qubit state can be created out of any
other qubit state via this operation. Within the one-way framework
this task is implemented via adaptive measurements, in a five
qubit cluster state~\cite{oneway}. The measurement pattern and
required adaptations are depicted in Fig.~\ref{fig:primitives} a).
Given that the input qubit was initially in the state
$\ket{\psi_{\text{in}}}$, after a execution of this protocol with
outcomes $\vec{r}$, the output qubit is left on $BP_{\vec{r}}\,
R(\phi_1,\phi_2,\phi_3)\ket{\psi_{\text{in}}}$. The by-products
$BP_{\vec{k}}$ of this computation are defined  as in~\eqref{BP}
with:
\begin{align*}
f_{5,Sig}(\vec{k}) &= k_3 k_2;\\
f_{5,X}(\vec{k}) &= k_4+k_2;\\
f_{5,Z}(\vec{k}) &= k_3+k_1.
\end{align*}

For the $CNOT$ gate a cluster of 15 qubits is necessary, but no
adaptations are required ~\cite{oneway}. The measurement pattern
defining the algorithm is shown in Fig.~\ref{fig:primitives} b).
The $CNOT$ acts on two qubits, called target and control, such
that if the control is in the state $\proj{1}$ the target qubit is
flipped and nothing happens otherwise, that is,
$CNOT_{ij}=\proj{0_i}\otimes\openone_j+\proj{1_i}\otimes X_j$. If
the initial control-target state in the qubits 1 and 9 is
$\ket{\chi_{\text{in}}}$, the outcome state in the qubits 7 and
15, after measurements with outcome $\vec{r}$, is
$\ket{\chi_{\text{out}}} = BP_{\vec{r}}\, CNOT
\ket{\chi_{\text{in}}}$. The by-products of this operation are
given by setting:
\begin{align*}
f_{7,X}(\vec{k})&=k_2 + k_3 + k_5 + k_6;\\
f_{15,X}(\vec{k})&=k_2 + k_3 + k_8 + k_{10} + k_{12} + k_{14};\\
f_{7,Z}(\vec{k})&= k_1 + k_3 + k_4 + k_5 + k_8 + k_9 + k_{11} + 1;\\
f_{15,Z}(\vec{k})&=k_9 + k_{11} + k_{13}.
\end{align*}
As no adaptations are necessary for this gate, we can ignore the
boolean function $f_{Sig}$.

Now consider that all measured qubits, for both protocols, are
under the influence of identical local environments $(p_i=p)$. As
before we consider the cases of phase-flip and white noise. The
noisy evolution of the fidelity for the rotation protocol,
$\mathcal{F}^{\text{w/pf}}_{R}$, can be assessed by
expressions~\eqref{fidelity} and~\eqref{fidaverage}. While for the
controlled-not gate, the dynamics of
$\mathcal{F}^{\text{w/pf}}_{CNOT}$ is obtained via
Eq.~\eqref{fidelityNA}, with no need of averaging over the
possible measurement outcomes. For the protocols do not need
measurements along the $z$ direction, if we have
$p_{xy}^{\text{w}}=p_{xy}^{\text{pf}}$, then the fidelity decay
under the two kinds of noise is exactly the same. That is, for
both noisy scenarios
$\mathcal{F}^{\text{pf}}_{R}=\mathcal{F}^{\text{w}}_{R}$ and
$\mathcal{F}^{\text{pf}}_{CNOT}=\mathcal{F}^{\text{w}}_{CNOT}$.
This is in stark contrast to the entanglement decay of the
resource states which will be generically different under the two
noise instances. See Fig.~\ref{fig:fidelities_rotation_cnot} for a
quantitative account of this fact. This once more shows that
entanglement dynamics is detached from the fidelity dynamics, and
as such a less entangled state can lead to higher (or equal)
quality computations.

%%%%%%%%%%%%%%%%% DJ section %%%%%%%%%%%%%%%%%%%%%%%%%%
%%%%%%%%%%%%%%%%%%%%%%%%%%%%%%%%%%%%%%%%%%%%%%%%%%%%%%
%%DJ circut model
%%%%%%%%%%%%%%%%%%%%%
\subsection{Deutsch-Jozsa(DJ) algorithm \label{DJalgorithm}}
 Let $f:\{0,1\}^N \mapsto \{0,1\}$ be an
unknown boolean function which can be either constant -- with all
entries giving the same answer--, or balanced -- with half of the
entries yielding $0$ and the other half $1$. What is the minimum
number of times one has to query an Oracle that implements $f$ to
discover the function's type? Classically, in general, one needs
at least $2^{N-1}+1$ queries. Quantum mechanically, a single query
via the DJ algorithm is sufficient~\cite{DJ92}. In the quantum
version, the Oracle applies a unitary $U_f$ on the $N$ qubits of
the input state plus an auxiliary qubit ($A$), as follows: $U_f
\ket{x}\ket{y}=\ket{x}\ket{y\oplus f(x)}$; where $x$ and $y$ are
the decimal representations of binary strings. The DJ algorithm
takes advantage of the superposition principle to evaluate all the
entries at once, and can be cast as the unitary $DJ = H^{\otimes
(N+1)} \, U_f \, H^{\otimes (N+1)}$, with
$H=1/\sqrt{2}\big(\proj{0}+\ket{0}\bra{1}+\ket{1}\bra{0}-\proj{1}\big)$
the  Hadamard gate. It is a simple calculation to deduce that
$DJ(\ket{0}^{\otimes N}\ket{1})$ leads to the outcome $\vec{0}$,
after measurement of the $N$ qubits in the computational basis, if
and only if $f$ is constant. This is clearly spelled out by the
state below, %%
\begin{equation}
\label{DJstate} DJ(\ket{0}^{\otimes N}\ket{1}) =H^{\otimes
N+1}\frac{1}{2^{N/2}} {\displaystyle\sum\limits_{i=0}^{2^{N}-1}}
(-1)^{f(i)}\left\vert i\right\rangle \left\vert -\right\rangle.
\end{equation}
If $f$ is constant the  final state gets only a global phase
$(-1)^{f(0)}$ in relation to the initial state (remember that
$H^2=\openone$).
%%%%%%%%%%%%
% figure 4
%%%%%%%%%%%%
\begin{figure} [t!]
\begin{center}
\includegraphics[width=\linewidth]{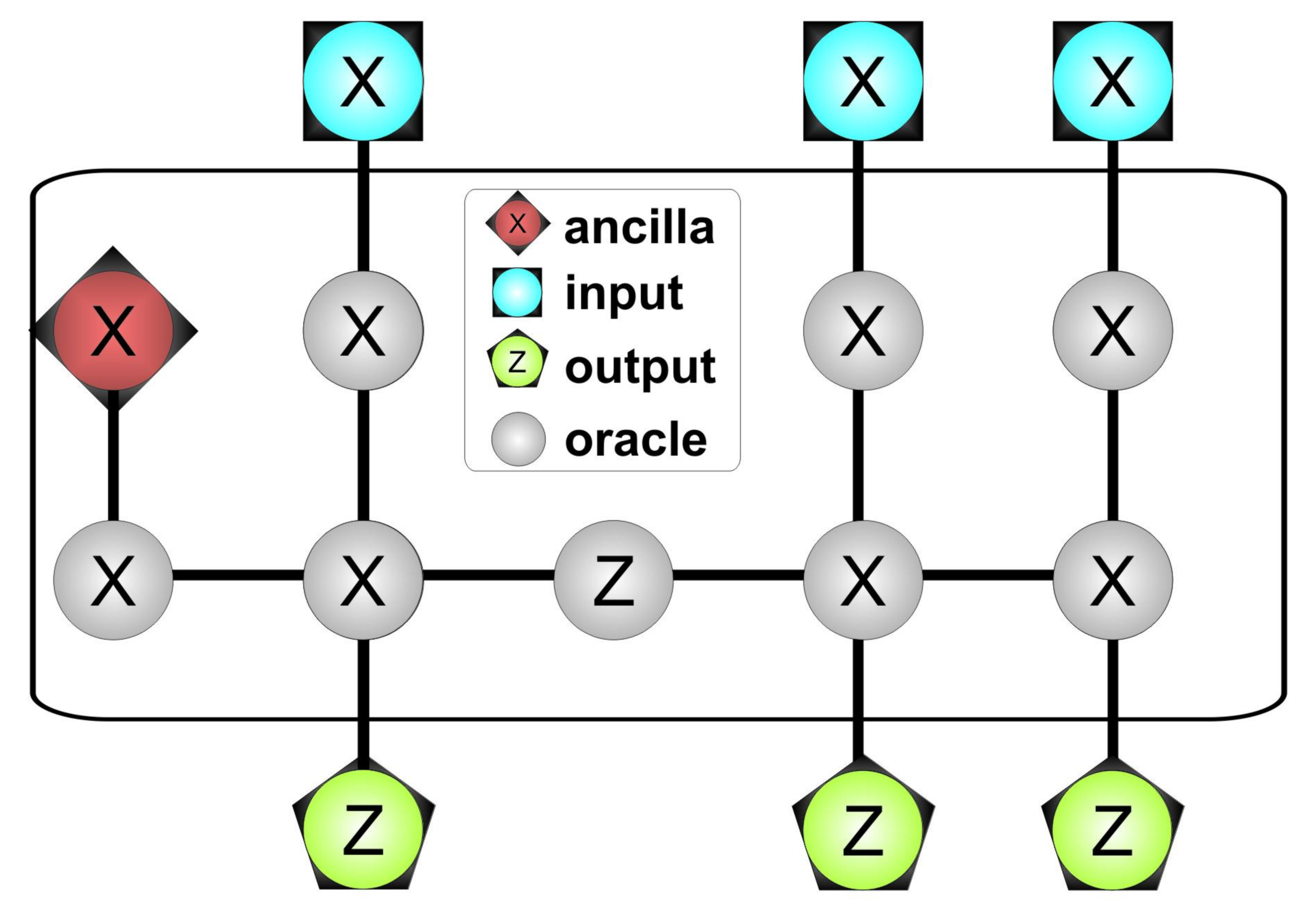}
\caption{(Color online) \textbf{Graph-state and measurement
pattern for the DJ protocol simulating the balanced function
$U_{f}= {\prod_{i=2}^{N-1}} CZ_{i,i+1}CNOT_{1,A}$}. All the qubits
are measured without adaptations, and so the
entangled graph-state can be replaced by a mixed state with only
classical correlations (see Sec.~(\ref{without_adaptations})). \label{fig:DJprotocolforn}}
\end{center}
\end{figure}
%%%%%%%%%%
%%%%%%%%%%
In its noiseless  implementation, the DJ is known to generically
create entanglement~\cite{EntangledDJ1998}. Note that along all
the protocol the auxiliary qubit is never entangled with the
$N$-qubit principal system. However, depending on the function $f$
that the Oracle implements, the $N$-qubit system can become
entangled. For a constant function $f$ the state is never
entangled as can be readily seen from (\ref{DJstate}). Despite of
that, for example, the balanced function $f: \{0,1\}^3 \rightarrow
\{0,1\}$ with truth table:
\begin{center}
\begin{tabular}{c|c}
\hline
$x$&$f(x)$\\ \hline
0 = 000 & 0\\
1 = 001 & 0\\
2 = 010 & 0\\
3 = 011 & 1\\
4 = 100 & 1\\
5 = 101 & 1\\
6 = 110 & 1\\
7 = 111 & 0
\end{tabular}
\end{center}
generates  an entangled state during the execution of the DJ
protocol \cite{separable multipartite states}. For three qubits
this balanced function can be performed  by the unitary operation
$U_{f}=CNOT_{1,A}CZ_{2,3}$, and can be easily generalized for $N$
qubits as $U_{f}= CNOT_{1,A}{\bigotimes_{i=2}^{N-1}} CZ_{i,i+1}$.
Note that the operation $U_{f}$ can be done within the one-way
framework by measurements without any adaptations (see
Fig.~(\ref{fig:DJprotocolforn})).

On the other hand, a noisy implementation of the DJ, still within
the circuit model, was analyzed in~\cite{Biham2004} and shown to
present a small but finite advantage over its classical
counterpart even for fully separable states. This indicates that
to unveil the role of the entanglement in the mixed DJ protocol
one has to split the generation and use of entanglement.
%%%%%%%%%%%%%%%%%%%%%%%%%
%%DJ one way
%%%%%%%%%%%%%%%%%%%%%%%%%
In the one-way setting the DJ problem can be posed as
follows~\cite{experimental DJ}: the Oracle prepares a graph state
that allows her to implement a function $f$ by local measurements.
She hands in to Neo (the user) a set of input qubits, and a set of
output qubits. By encoding (through measurements) a certain value
$x$ into the input qubits, Neo can read out $f(x)$ in the output
qubits. As before, an appropriate choice of measurements allows
Neo to discover whether the function is constant or balanced in a
single run of the protocol.

In the example shown in Fig.~(\ref{fig:DJprotocolforn}) neither
the implementation of $f$ nor the measurements by Neo require
adaptations. It is thus possible for the Oracle to interchange the
NAMs in a entangled graph state with a \emph{mixed state without
any entanglement}, and even though Neo can decide whether $f$ is
constant or balanced in a single query. To design such classically
correlated state the Oracle proceeds as follows: \emph{i)} think
of the original cluster state needed to encode the desired
function for any input; \emph{ii)} apply to it an hypothetical  noise of the
type in (\ref{maps2fixedpoles}), but protecting the measurement
outcomes by rotating the qubits to a convenient basis. In this step the Oracle can only protect a single instance of input states, and she does that for the $H^{\otimes N}\ket{\vec{0}}$,  \emph{iii)}
finally, she evaluates the stationary state of the decoherence process,
and rotates the qubits back to their original basis. The resulting
(theoretical) state can be effectively prepared by the Oracle only
by classical means, and it is ready to be hand in to Neo.   If now
Neo performs the correct set of measurements on the input qubits, the
output qubits will encode whether the function is constant or
balanced. It is interesting to notice that despite the fact the
state bears only classical correlations, the quantum possibility
of measuring on different basis entails advantages over a fully
classical implementation. In fact, many boolean functions,
that in a circuit model generate entangled states, can be
decomposed into combinations of $CZ$'s, $CNOT$'s, $H$'s and possible
relabelling of the qubits, transformations that can be attained
within the one-way model without adaptations. As such, they can be simulated by a classically correlated resource state. These ideas, therefore, extend to many algorithms, for instance to the Simon's~\cite{simon} algorithm, as long as the function evaluated by the Oracle does not require adaptations. An interesting question, that we leave open, is to determine the fraction of boolean functions that can be evaluated without adaptations.

%%%%%%%%%%%%%%%%%%%%%%%%%
%%Ancilla Driven Computation
%%%%%%%%%%%%%%%%%%%%%%%%%
\subsection{Ancilla-Driven Quantum Computation \label{ancilla}}

\begin{figure} [t!]
\begin{center}
\includegraphics[width=\linewidth]{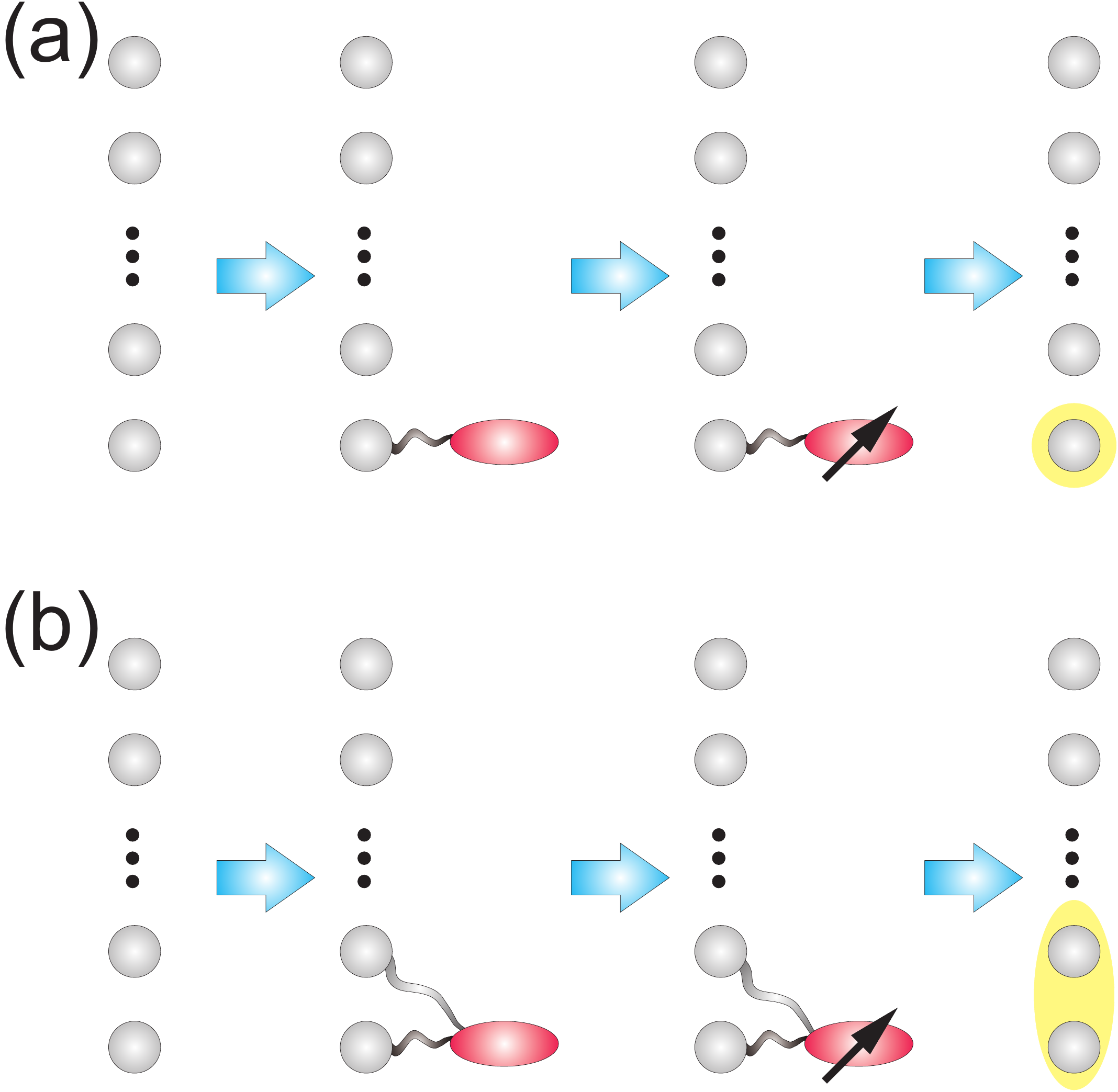}
\caption{(Color online) \textbf{(a) Single qubit rotation}. After
the interaction $E=H_{a}\,H_{r_{i}}\,CZ_{ar_{i}}$ between the
$i$-th qubit on the register and the ancilla, the latter is
measured in the $ 1 / \sqrt{2} \left( \left\vert 0\right\rangle
\pm e^{i\phi} \left\vert 1\right\rangle \right)$ basis. As a
result the transformation $X^{k}H R_{\hat{z}}(\phi)$ is applied to
the qubit in the register (with $k$  the classical outcome of the
measurement). \textbf{(b) Two qubit entangling gate}. After the
interaction the ancilla is measured in the  basis $\left\{
\left\vert 0\right\rangle , \left\vert 1\right\rangle\right\}$ and
as a result the two qubits on the register undergo the
transformation  $(X_{i}^{k} H_{i}\otimes H_{j}) CZ_{ij}$}
\end{center}
\label{fig:ancillacomputation}
\end{figure}

 The idea used in Sec.~\ref{onewaymodelunderdecoherence} can be readily
applied to the ancilla-driven model of quantum computation
\cite{ancilladriven}. In this model all the
necessary unitary transformations to be realized in a quantum
register are done through measurements in an ancillary qubit
coupled to the register. A coupling interaction given simply by
$E=H_{a}\,H_{r_{i}}\,CZ_{ar_{i}}$ is sufficient to allow for universal
quantum computation. The index $a$ represents the ancilla,
and $r_{i}$ the $i$-th qubit of the quantum register. After the
interaction the ancilla is measured in the same basis~(\ref{base})
of the usual one-way model (see
Fig.~(\ref{fig:ancillacomputation}) for details). %%
%%%%%%%%%%%%
% figure 5
%%%%%%%%%%%%
%%%%%%%%%%
%%%%%%%%%%
The effects of an inaccurate measurement (the projection direction
is deviated from the ideal one) in the fidelity of the
ancilla-driven quantum computation was analysed in
\cite{Tomoyuki2}. There it was shown that for a faulty
measurement, the more the qubit being coupled to the ancilla is
entangled with the rest of the qubits in the register, the smaller
is the mean fidelity of the operation. The relation is given by
\begin {equation}
\overline{\mathcal{F}} \le 1- S_L \sin ^{2}  \frac {\epsilon}{2},
\label{measurementfidelityrelation}
\end {equation}
where the entanglement is quantified by $S_L=2 \left[1- \tr
\left(\rho_{r_{i}}^2 \right) \right]$, the normalized linear entropy of the
reduced density matrix $\rho_{r_{i}}$, and $\epsilon$ quantifies
the angular deviation from the ideal measurement. Here, we will derive an analogous
expression, but instead of an inaccurate measurement
we consider that the ancilla qubit is undergoing one of
the decoherence processes described by the
map~(\ref{generalchannel}). In this scenario we arrive
at the relation
\begin{equation}
\overline{\mathcal{F}} \le 1- p\, S_L, \label{decofidelityrelation}
\end{equation}
which is fully equivalent to (\ref{measurementfidelityrelation}).
For a given amount $S_L$ of entanglement we can thus estimate how small
should be the decoherence parameter $p$ such that the computation
fidelity is not below a given threshold.

The derivation of~\eqref{decofidelityrelation} follows the same
ideas as in Ref.~\cite{Tomoyuki}. Consider the register state
$\left\vert \psi_{r}\right\rangle$ is given by %%
\begin{equation}
\left\vert \psi_{r}\right\rangle=\alpha\left\vert
0\right\rangle\left\vert \eta_{0}\right\rangle +\beta\left\vert
1\right\rangle\left\vert \eta_{1}\right\rangle.
\label{registerstate}
\end{equation}
After the interaction of a register-qubit with the ancilla
(initialized in $\left\vert +\right\rangle$) the register-ancilla
state $\left\vert \phi_{ar}\right\rangle$ is given by (see Fig.
\ref{fig:ancillacomputation} a)
\begin{eqnarray}
\left\vert \phi_{ar}\right\rangle=\alpha\left\vert
0\right\rangle_{a}\left\vert +\right\rangle\left\vert
\eta_{0}\right\rangle +\beta\left\vert
1\right\rangle_{a}\left\vert -\right\rangle\left\vert
\eta_{1}\right\rangle \\
=\frac{1}{\sqrt{2}} \left( \left\vert
M_{0}\right\rangle_{a}\left\vert A_{0}\right\rangle + \left\vert
M_{1}\right\rangle_{a}\left\vert A_{1}\right\rangle \right),
\label{registerstateafterinteraction}
\end{eqnarray}
where $\left\vert A_{k}\right\rangle=\alpha\left\vert
+\right\rangle\left\vert \eta_{0}\right\rangle +(-1)^k
e^{i\phi}\beta\left\vert -\right\rangle\left\vert
\eta_{1}\right\rangle$. From the results of the Sec.
\ref{onewaymodelunderdecoherence} we know that the effect of a map
(\ref{generalchannel}) on the ancilla  is to mix the answers of
the computation. It is straightforward to calculate the mean
fidelity $\overline{\mathcal{F}}=
\mathcal{Z}_{0}\mathcal{F}(0)+\mathcal{Z}_{1}\mathcal{F}(1)$ (with
$\mathcal{Z}_{k}$  the probability of measuring $\left\vert
M_{k}\right\rangle$ with a associated fidelity $\mathcal{F}_{k}$)
given by
\begin{equation}
\overline{\mathcal{F}}=1-p \left( 1- \xi^2 \right), \label{fidelityancilla}
\end{equation}
where $\xi=\tr \left( Z_{k}\rho_{k} \right)$. Since $S_L \le 1-
[\tr\left( Z_{i}\rho_{i} \right)]^2$ we  immediately obtain
relation (\ref{decofidelityrelation}). The same result holds for
the situation where the ancilla is coupled to two qubits of the
register state (see Fig. \ref{fig:ancillacomputation}(b)).

%%%%%%%%%%%%%%%%%%%%%%%%%%%%%%%%%%%%%%%%%
%%%%%%%%%%% Conclusions %%%%%%%%%%%%%%%
%%%%%%%%%%%%%%%%%%%%%%%%%%%%%%%%%%%%%%%

\section{Conclusion}
\label{conclusion}

In this paper we have shown that the effect of very general local
decoherence maps on the measurement bases employed by the one-way
model is to mix the two measurement directions with a certain
weight $p$, which characterizes the map. With this observation the
fidelity of any computation within the one-way model can be
readily obtained. This allowed us to conclude that the impact of
the noise on the entanglement in the resource state  is not
generically related to loss of computation quality.

Even for the simplest one-way protocol, the remote state
preparation (Sec.~\ref{RSPqubits}), a state with  more
entanglement (or discord) does not necessarily yield a higher
quality computation. In other words, robust entanglement does not
mean robust one-way quantum computations. Rather on the contrary,
for the ancilla-driven measurement based computation model
(Sec.~\ref{ancilla}), the fidelity sensitivity to decoherence is
bigger the higher is the entanglement in the resource state.

Even more surprisingly, the framework developed here made clear
that the parts of algorithms that do not require adaptations can
be replaced by a classically correlated resource state. This
implies, for example, that some instances of the Deutsch-Josza
algorithm can be realized in the one-way paradigm without any
entanglement (Sec.~\ref{DJalgorithm}).

Thus, if entanglement (discord) cannot be assigned as the
signature of efficient noisy quantum computations, which other
quantities may assume this role? For the RSP protocol, the amount
of non-classicality in the resource state (quantified by the
minimum entanglement potential) seems to also point out the
computation quality (Sec.~\ref{RSPqubits}). We believe that would
be interesting to extend or falsify this connection between
non-classicality and computation fidelity to more general cases.

All these results show that entanglement may not be the most
important resource for the quality of noisy quantum computations.
The mere use of a quantum logic, in a mixed state scenario, seems
to entail considerable gain over classical computations. This has
obvious implications to experimental implementation of quantum
information processes, for it relaxes the required isolation of
the quantum system from its environment. In fact, this can shed
some light on the functioning of biological systems, which,
despite of being strongly influenced by the surroundings, are
mesoscopic systems that may profit from  their quantum
nature~\cite{bio}.

\bigskip

\textbf{Acknowledgements} We would like to acknowledge Tomoyuki
Morimae for comments and for pointing out
references~\cite{Tomoyuki,Tomoyuki2}. R. C. was funded by PROBRAL
CAPES/DAAD, Faperj and Q-Essence project. F. de M. was supported
by Alexander von Humboldt Foundation, and the Belgian
Interuniversity Attraction Poles Programme P6/02.


\begin{thebibliography}{99}
\bibitem {qc}M. A. Nielsen, and I. L. Chuang, \emph{Quantum Computation and
Quantum Information} (Cambridge, Cambridge, 2000).
\bibitem {Jozsa2003} R. Jozsa, and N. Linden, Proc. Royal Soc. London {\bf 459}, 2011 (2003).
\bibitem {Vidal2003}G. Vidal, Phys. Rev. Lett \textbf{91}, 147902
(2003).
\bibitem {KnillGottesman1999}D. Gottesman, and I. Chuang, Nature \textbf{390}, 402 (1999).
\bibitem {Knill1998}E. Knill, and R. Laflamme, Phys. Rev. Lett. {\bf 81}, 5672 (1998).
\bibitem {DattaThesis}A. Datta, \emph{Studies on the Role of Entanglement in Mixed-state Quantum Computation}, PhD thesis, 2008, arXiv:0807.4490.
\bibitem {Jones2001}J. Jones, Prog. NMR Spectrosc. {\bf 38}, 325 (2001).
\bibitem {Braunstein98}S. L. Braunstein et al., Phys. Rev. Lett. {\bf 83}, 1054 (1999).
\bibitem {classical}M. A. Nielsen, E. Knill, and R. Laflamme, Nature {\bf 396}, 52 (1998). L. M. K. Vandersypen et al., Nature {\bf 414}, 883 (2001).
\bibitem {oneway}R. Raussendorf, and H. J. Briegel, Phys. Rev. Lett. {\bf 86}, 5188 (2001);
R. Raussendorf, D. E. Browne, and H. J. Briegel, Phys. Rev. A {\bf
68}, 022312 (2003).
\bibitem {TooEntangled} M. J. Bremner, C. Mora, and A. Winter, Phys. Rev. Lett. {\bf 102}, 190502
(2009); D. Gross, S. T. Flammia, and J. Eisert, Phys. Rev. Lett.
{\bf 102}, 190501 (2009).
\bibitem {DJ92} D. Deutsch, and R. Jozsa, Proc. R. Soc.  A {\bf 439}, 553558 (1992).
\bibitem {EntangledDJ1998} D. Collins, K. W. Kim, and W. C. Holton,
Phys. Rev. A \textbf{58}, R1633 (1998).
\bibitem{ancilladriven} J. Anders et al., Phys. Rev. A  \textbf{82}, 020301(R) (2010).
\bibitem {RemStatePrep}C. H. Bennett et al., Phys. Rev. Lett. \textbf{87}, 077902 (2001).
\bibitem {SingletFraction} M. Horodecki, P. Horodecki, and R. Horodecki, Phys. Rev. A\textbf{60}, 1888 (1999).
\bibitem {Weinstein}Y. S. Weinstein, Phys. Rev. A \textbf{79}, 052325 (2009).
\bibitem {Chaves} R. Chaves, and L. Davidovich, Phys. Rev. A \textbf{88}, 052308 (2009).
\bibitem {Tomoyuki}T. Morimae, Phys. Rev. A \textbf{81}, 060307(R) (2010).
\bibitem {Tomoyuki2}T. Morimae, and J. Kahn, Phys. Rev. A \textbf{82}, 052314 (2010).
\bibitem {preskill} J. Preskill, Proc. R. Soc. Lond.   \textbf{454},  385 (1998).
\bibitem {dur05}M. Hein, W. D{\"u}r, and H. J. Briegel, Phys. Rev. A {\bf 71}, 032350 (2005).
\bibitem {GraphStatesEntanglement}D. Cavalcanti et al., Phys. Rev. Lett. {\bf 103}, 030502 (2009); O. G{\"u}hne, F. Bodoky, and M. Blaauboer, Phys. Rev. A \textbf{78}, 060301(R) (2008); L. Aolita et al., Phys. Rev. A \textbf{82}, 032317 (2010).
\bibitem {terra01}R. C. Drummond, and M. O. Cunha, J. Phys. A {\bf 42}, 285308 (2009).
\bibitem{montanaro} A. Montanaro, arXiv:1001.0018 (2010).
\bibitem {wooters98}W. K. Wootters, Phys. Rev. Lett. {\bf 80}, 2245 (1998).
\bibitem {negativityvidal}G. Vidal and R. F. Werner, Phys. Rev. A {\bf 65}, 032314 (2002).
\bibitem{nota} Bellow the $2/3$ threshold, depicted by the horizontal dashed line in Fig.~\label{fig:counterExampleentanglement}, a model of classical local-hidden variables is sufficient to reproduce all the correlations presented by the resource state, and thus also to achieve the obtained computation fidelity.
\bibitem {Ollivier2001} H. Ollivier, and W. H. Zurek, Phys. Rev. Lett. {\bf 88}, 017901 (2001).
\bibitem {luo08} S. Luo, Phys. Rev. A {\bf 77}, 042303 (2008).
\bibitem {MEP}M. Piani, S. Gharibian, G. Adesso, J. Calsamiglia, P. Horodecki, A. Winter, arXiv:1103.4032v1 (2011).
\bibitem {separable multipartite states} P. Jorrand, and M. Mhalla, IJFCS \textbf{14}, 797 (2003).
\bibitem {Biham2004}E. Biham et al., Theo. Comp. Sci. {\bf 320}, 15 (2004).
\bibitem {experimental DJ}M. S. Tame et al., Phys. Rev. Lett. {\bf 98}, 140501 (2007);
G. Vallone et al., Phys. Rev. A {\bf 81}, 050302 (2010); M. S. Tame, and  M. S. Kim, arXiv:1003.4974v2 (2010).

\bibitem {simon} Simon, D. R., SIAM J. Comput. {\bf 26}, 1474 (1997).
\bibitem{bio} G. Engel et al., Nature {\bf 446}, 782 (2007); T. Scholak et al. Phys. Rev. E {\bf 83}, 021912 (2011).

\end{thebibliography}
\end{document}